\documentclass[letterpaper, 10 pt, conference]{ieeeconf} % Comment this line out if you need a4paper

\IEEEoverridecommandlockouts                              % This command is only needed if 
\usepackage{amsmath} % assumes amsmath package installed
\usepackage{amssymb}  % assumes amsmath package installed

\usepackage{comment}
\usepackage{acronym}
\usepackage{tikz,pgfplots} % Allows drawing vector graphics
\usetikzlibrary{positioning, arrows.meta, calc, fadings, shadows}
\usepgfplotslibrary{colorbrewer}
\pgfplotsset{compat=1.18}
\usepackage{macros}
\usepackage{physics}
\usepackage{siunitx}
\usepackage{colorbar}
\usepackage{hyperref}
\usepackage{cleveref}
\usepackage{makecell}
\usepackage{pifont}
\usepackage{cite}

\usepackage[font=footnotesize]{caption}
\usepackage{subcaption}

\usepackage{algorithm2e}
\SetKwComment{Comment}{/* }{ */}
\RestyleAlgo{ruled}

\newcommand{\R}{\mathbb{R}}
\def\always{\square}
\def\event{\lozenge}
\def\transp{\mathsf{T}}

\newcommand\scalemath[2]{\scalebox{#1}{\mbox{\ensuremath{\displaystyle #2}}}}

\newtheorem{proposition}{Proposition}

\title{\LARGE \bf
Learning-enabled Parameter Synthesis for Nonlinear Systems \\from Signal Temporal Logic 
}

\author{Alex Beaudin$^{1,*}$, Hanna Krasowski$^{1,*}$, Eric Palanques-Tost$^{2}$, Calin Belta$^{3}$, and Murat Arcak$^{1}$% <-this % stops a space
\thanks{*Equal contribution}% <-this % stops a space
\thanks{$^{1}$ University of California, Berkeley
        {\tt\small \{alex\_beaudin, krasowski, arcak\}@berkeley.edu}}%
\thanks{$^{2}$ Boston University
        {\tt\small ericpt@bu.edu}}%
\thanks{$^{3}$ University of Maryland, College Park
        {\tt\small cbelta@umd.edu}}%
\thanks{This work was funded in part by the Air Force Office of Scientific Research grant FA5590-23-1-0529. Alex Beaudin is partially supported by a Fonds de Recherche du Qu\'ebec Doctoral Scholarship.}
}

\begin{document}

\maketitle
\thispagestyle{empty}
\pagestyle{empty}

%%%%%%%%%%%%%%%%%%%%%%%%%%%%%%%%%%%%%%%%%%%%%%%%%%%%%%%%%%%%%%%%%%%%%%%%%%%%%%%%
\begin{abstract}
Signal Temporal Logic (STL) is increasingly used to describe interpretable objectives and constraints for optimal control and learning methods, especially when no target time series data is available. In this work, we propose to synthesize parameters for nonlinear systems that robustly satisfy continuous-time STL specifications for uncertain initial conditions. To this end, we use gradient-based optimization along with set-based reachability verification to efficiently learn in high-dimensional parameter spaces while providing provable satisfaction guarantees for the optimized parameters. We demonstrate the effectiveness and scalability of our method on three systems with up to 18 parameter dimensions.
\end{abstract}

%%%%%%%%%%%%%%%%%%%%%%%%%%%%%%%%%%%%%%%%%%%%%%%%%%%%%%%%%%%%%%%%%%%%%%%%%%%%%%%%
\section{Introduction}
\begin{figure*}[tb]
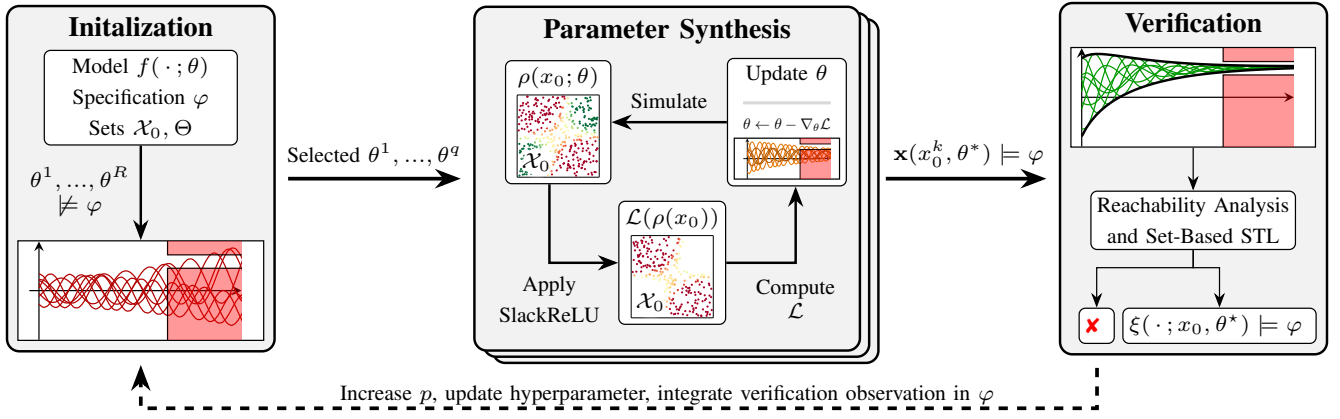

    \centering
    \include{figures/headfigure}
    \vspace{-2em}
    \caption{Left to right: 1) finding parameters that satisfy the specification $\varphi$ is challenging, due to the non-smooth learning landscape as illustrated in Fig.~\ref{fig:landscape}. 2) We propose a gradient-based optimization that yields parameterizations of the system $f$ such that the specification is satisfied for all finitely many samples. This is illustrated by a scatterplot over the initial set $\mathcal{X}_0$ where the color indicates the robustness of the trajectory starting at that initial condition. 3) To obtain continuous domain guarantees, we compute reachable sets and verify them with respect to the specification. }
    \label{fig:headfigure}
\end{figure*}

Temporal logic is a widespread family of unambiguous and interpretable formal languages.
Traditionally, it has been employed in verification with an emphasis on discrete space and time systems.  
\ac{stl}~\cite{maler2004, Fainekos2009} describes continuous time and space requirements (specifications). 
Beyond verification, \ac{stl} has been used for controller synthesis~\cite{Meng2023, SrinivasanCoogan2021-TL2CBF-TRO}, monitoring \cite{pek2023spatial, deshmukh2017robust}, reinforcement learning \cite{Li2017, Jiang_Bharadwaj_Wu_Shah_Topcu_Stone_2021}, and parameter synthesis \cite{Krasowski2025a, PalanquesTost2025}.
In this paper, we present an approach that combines a learning-based algorithm for parameter synthesis from STL with model-based verification (see Fig.~\ref{fig:headfigure}). This approach obtains continuous-time satisfaction guarantees for general nonlinear systems with bounded initial conditions by splitting the synthesis into a learning step and a verification step. 
The learning-based algorithm leverages residual \acp{node} \cite{chen2018neural}, counter examples, and \ac{stl}-informed parameter initialization to mitigate the challenges of learning with \ac{stl}-based loss functions. 
We show that our approach scales to high state and parameter dimensions on a variety of nonlinear dynamical models. 
% The main contributions are:
% \begin{itemize}
%     \item We propose a scalable learning algorithm to identify satisfying parameters of continuous-time nonlinear systems from \ac{stl} specifications;
%     \item We extend the learning algorithm with a model-based a posteriori verification, rendering the same satisfaction assurance as symbolic synthesis methods;
%     \item We demonstrate the proposed algorithm on three benchmarks with varying specification complexity and parameter dimensionality.
% \end{itemize}
Our main contributions are:
\begin{itemize}
    \item we propose a scalable algorithm to identify parameters of continuous-time nonlinear systems from \ac{stl} specifications,
    \item we extend the algorithm with a posteriori verification, yielding the same satisfaction assurance as symbolic synthesis methods, and
    \item we demonstrate the proposed algorithm with varying specification complexity and parameter dimensionality.
\end{itemize}

\section{Related Work}

\textbf{Parameter synthesis. } There are two dominant threads for parameter synthesis from \ac{stl} specifications: symbolic methods and simulation-based methods.
Symbolic methods rely on analytical representations of system dynamics to derive parameter sets with formal satisfaction guarantees. 
For instance, \cite{dang15,sapo} proposes a recursive set propagation and Bernstein polynomial relaxations to compute parameter regions that provably satisfy the specification. 
Related approaches exploit domain-specific structures to enable computationally tractable synthesis~\cite{Bartocci2013, demko2016high}. 
While these methods provide strong guarantees, they make strong assumptions on the system dynamics, and the set propagation technique scales poorly with the number of parameters.

Simulation-based methods empirically verify satisfaction by evaluating the specification on a finite set of trajectories. 
Rather than guaranteeing satisfaction, they optimize parameters to satisfy the specification only on the sampled, usually discrete-time, trajectories \cite{PalanquesTost2025, Krasowski2025a, bortolussi18}. %, bartocci2015system, donze2010breach}.
% In this sense, the problem turns into a standard optimization task over simulated executions, rather than a global verification problem, which allows relaxing the constraints on the system. 
Earlier work adopts Bayesian optimization with Gaussian process surrogates to guide parameter exploration~\cite{bortolussi18}. 
Recent approaches leverage gradient-based methods~\cite{PalanquesTost2025, Krasowski2025a}, improving scalability in parameter dimensions. 
However, they do not provide guarantees, and the non-smoothness of the robustness function leads to unstable optimization.
In summary, existing approaches prioritize either formal guarantees or scalability. In contrast, our neuro-symbolic method leverages learning-based scalability alongside tractable a posteriori verification.
%to obtain continuous time guarantees.

% \textcolor{gray}{[Our xxx approach relates to learning-based and symbolic synthesis of system parameters. We also discuss relevant verification approaches that allow for establishing symbolic synthesis level satisfaction guarantees.]}

\textbf{System verification. }
For continuous-time nonlinear systems, set-based reachability analysis provides formal safety guarantees and scales to high-dimensional systems \cite{althoff2021set}. 
Further, reachability analysis has been extended to verify richer temporal logic properties through set-based \ac{stl} model checking, including reachset temporal logic evaluation~\cite{roehm2016stl}, incremental verification schemes~\cite{lercher24}, and system class-specific approaches for linear or structured systems~\cite{kochdumper2024}. 
A complementary direction to \ac{stl} model checking 
%include Hamilton–Jacobi reachability~\cite{Mitchell2005,chen2018signal} and 
is a mixed-integer programming formulation \cite{Raman2014}, which combines synthesis and verification, but is more computationally demanding. %for reasoning about temporal specifications. 
In this work, we leverage set-based reachability analysis to achieve continuous-time specification compliance of a given parameter vector for a set of initial conditions. % given a parameter vector that empirically satisfies the specification.

\section{Preliminaries and Problem Formulation}

We denote sets by calligraphic letters.
Curly brackets denote unordered sets, while parentheses denote ordered sets, e.g., $( x_i )_{i=1}^L$ is an ordered set of the states $x_i \in \mathbb{R}^n$, where $i$ is an index.
%Functions (right now functions and variables are overlapping)
%A Time discretion signal in bold font -- need to change alg notation/

\subsection{Parametric Models} 
We consider dynamic parametric models described by \acp{ode} %{\tiny \color{white} \ac{ode}}
\begin{equation}\label{eq:ode}
    \begin{aligned}
        \dot{x} &= f(x; \theta), \quad x(0) = x_{0},
    \end{aligned}
\end{equation}
where $\theta$ is the parameter vector, and $x_{0} \in \mathcal{X}_0 \subset \R^n$ is an initial condition for the system.
We denote the unique solution, or trace, of the initial value problem in \eqref{eq:ode} up to time $T$ by $\xi(\,\cdot \,;\,x_0, \theta): [0, T] \to \R^n$ and, its time-discretized analog by $\mathbf{x}(\,x_0, \theta) = [\xi(t_1; x_0, \theta),\dotsc, \xi(t_m; x_0, \theta)]$ for time points $t_1, \dotsc, t_m$, where $0 \leq t_1 < t_2 < \cdots < t_m \leq T$.
The forward reachable set for the time interval $[0,T]$ is
\begin{equation}\label{eq:reachableset}
    \mathcal{R}_{[0,T]} (\theta) = \{ (\xi(t;x_0, \theta), t) \;|\; t \in [0, T], \; x_0 \in \mathcal{X}_0\}.
\end{equation}

\subsection{Signal Temporal Logic}
We use temporal logic specifications defined by \ac{stl} formulas~\cite{maler2004} to capture system requirements.
Specifically, an \ac{stl} formula has three components: 1) predicates $\mu := g(x) \ge 0$ with $g: \R^{n} \to \R$; 2) standard Boolean operators like negation ($\lnot$), conjunction ($\land$), and implication ($\Rightarrow$); 
and 3) temporal operators, such as always $\square$ and eventually $\lozenge$.
Formally, an \ac{stl} formula has the syntax
\begin{equation}\label{eq:stl}
    \varphi := \mu \, | \, \varphi_1 \lor \varphi_2 \,|\, \varphi_1 \land \varphi_2 \,|\, \lnot \varphi \,|\, \square_{I} \varphi \,|\, \lozenge_I \varphi
                   \,|\, \varphi_1 \Rightarrow \varphi_2.
\end{equation}
For example, the \ac{stl} formula given by $\square_{[ t_1, t_2 ]} \varphi$ is true if $\varphi$ holds for all time in the interval $I = [t_1, t_2]$.
Similarly, $\lozenge_{[t_1, t_2]} (\varphi_1 \land \varphi_2)$ is true if there is a time $t\in [t_{1}, t_{2}]$ for which $\varphi_1$ and $\varphi_2$ are simultaneously true.
% We note that the syntax above contains redundant clauses, but defines all the symbols we use in this article.
We note that the syntax above defines all the symbols we use in this article at the expense of redundancy.
%\eric{May be clearer to define the predicates and then the temporal operators separately? Right now this formulation forces you to use $\square_{I}$ with the undefined symbol $I$. It also never defines $\mu$. It allows predicates $\mu$ without temporal operators.}

\ac{stl} specifications are defined over functions of time, or traces, usually generated by a dynamical system.
In this work, we consider both continuous and discrete-time \ac{stl} semantics.
For discrete-time semantics \cite[Def. 22]{Fainekos2009}, if a trace $\vb{x} \in \mathbb{R}^{n \times m}$, where $m$ is the number of time points,
is such that $\varphi(\vb{x})$ is true, then we write $\vb{x} \models \varphi$ and say $\vb{x}$  \textit{satisfies} $\varphi$.
Intuitively, these Boolean semantics denote whether a signal either does or does not satisfy the specification $\varphi$.
Additionally, quantitative semantics describe the degree of satisfaction through the robustness function $\rho(\vb{x}, \varphi)$, whose sign agrees with the Boolean semantics~\cite{Fainekos2009}, i.e., $\vb{x} \models \varphi \iff \rho(\vb{x} , \varphi) \geq 0$.
When the specification holds for the discrete-time traces of $K < \infty$ initial conditions (usually sampled randomly), we say it is \textit{empirically satisfied} and write
\begin{equation}\label{eq:discretetimeproblem}
    \vb{x} (x^k_{0}, \theta) \models \varphi  \ \text{ for all }  k \in \{1, ..., K\}.
\end{equation}
Analogously, for continuous-time traces \cite[Def. 5]{Fainekos2009}, the semantics are defined using $\xi$ instead of $\mathbf{x}$.

\subsection{Problem Statement}\label{sec:problem}

Given a parametric model \eqref{eq:ode} with a bounded set of initial conditions $\mathcal{X}_0$, a time horizon $[0,T]$, and an \ac{stl} specification $\varphi$ describing the desired system behavior, the problem is to identify system parameters $\theta^{\star} \in \Theta$ such that 
\begin{equation}\label{eq:conttimeproblem}
\xi(\,\cdot\, ;x_0, \theta^\star)\models \varphi  \text{ for all } x_0 \in \mathcal{X}_0 .
\end{equation}
%Note that \eqref{eq:conttimeproblem} demands continuous-time satisfaction over $\mathcal{X}_0$.
%To make the problem well-posed, we assume that there exists $\theta^\star \in \Theta$ such that $\mathcal{X}(\theta^\star) \models \varphi$.
%Note that this problem is only reasonably well posed if it is likely that such a $\theta^{\star}$ exists.

\section{Method}
\label{sec:method}
Solving this problem solely with symbolic methods provides continuous-time satisfaction, but it is not computationally tractable for high-dimensional nonlinear systems with many unknown parameters. 
Our proposed solution is to find a solution to~\eqref{eq:discretetimeproblem} with a gradient-based algorithm and afterward verify~\eqref{eq:conttimeproblem} with model-based reachability analysis. We illustrate the approach in Fig.~\ref{fig:headfigure} and structure it into three phases: initialization, parameter synthesis with \ac{sgd}, and continuous-domain verification.

\textbf{Initialization. }
First, we search for multiple parameter initializations, as well as maintain a counter-example buffer to prioritize challenging initial conditions. 
Specifically, we sample $R$ parameter candidates from the parameter space $\Theta$ and evaluate their robustness across $P$ samples $\{ x_0^p\}_{p=1}^{P}$ drawn uniformly from $\mathcal{X}_0$. We take the top $q$ candidates, ranked by worst-case robustness over $\{x_0^p\}_{p=1}^{P}$.
%, which we optimize in parallel using JAX~\cite{jax2018}. 

\begin{algorithm}[tb]
    \caption{Parameter Synthesis with SGD}\label{alg:sgd}
    \KwData{Initial set $\mathcal{X}_0$, time horizon $T$, parameter domain $\Theta$, specification $\varphi$, number of training epochs $E$, hyperparameters $m, R, P, q, N, M,\lambda, \alpha, \eta$, and
    test set $X_\mathrm{test}$.}
    \KwResult{$\theta \in \Theta$ such that $\vb{x} (x^k_{0}; \theta) \models \varphi$}
    Initialize \acp{node} $\zeta$ and slack parameters $\nu$;\\
    Initialize individual counter-example buffers $\mathcal{B}^q = \emptyset$;  \\
    $\Theta_q \gets \mathtt{get\_best\_params}(R, q, \varphi, \mathcal{T}(\Theta), P, U(\mathcal{X}_0))$; \\
    \For{$e \leq E$}{
    \For{$\theta^{i} \in \Theta_q$}{
        $x_0^k \sim \mathcal{D}(\mathcal{X}_0) \, \text{ for all }  \, k = \{1, ..., N\}$\;
        $x_b^j \sim U(\mathcal{B}^q_i) \, \text{ for all } \, j = \{1, ..., M\}$\;
        $\mathcal{X} \gets \{x_0^k\}_{k=1}^N \cup \{x_b^j\}_{j=1}^M $\;
        $L \gets \mathcal{L}(\theta^i, \nu^i, \alpha, \mathcal{X}, \varphi) + \lambda \|\zeta^i\|_2^2$ \;
        $\theta^i \gets \theta^i - \eta \grad_{\theta^i} L$, using AdaBelief\;
        $\zeta^i \gets \zeta^i - \eta \grad_{\zeta^i} L$, using AdaBelief\;
        $\nu^i \gets \nu^i - \eta \grad_{\nu^i} L$, using AdaBelief\;
        $\mathcal{B}^q_i \gets \{(x_0)^k \in \mathcal{X}_0 \,|\,\vb{x} (x^k_{0}, \theta) \not\models \varphi  \} \cup \mathcal{B}^q_i$\;
        $e \gets e+1$
        }
    }
    $\theta^* \gets \argmax_{\theta \in \{\theta^i\}} \min_{x_0^k \in X_\mathrm{test}} \rho(\vb{x}(x_0^k; \theta), \varphi)$;
\end{algorithm}

\textbf{Parameter Synthesis with SGD. }
We summarize our gradient-based optimization in Alg~\ref{alg:sgd}. We train $q$ parameter candidates in parallel with individual counter-example buffers. 
At each training iteration, we sample $N$ initial conditions uniformly from $\mathcal{X}_0$, and uniformly from the boundary of $\mathcal{X}_0$, since these points often correspond to extreme trajectories. We denoted this distribution by $\mathcal{D}$. Re-sampling $\{x_{0}^k\}_{k=1}^{N}$ each training epoch renders the objective \eqref{eq:lossobjective} an approximation of the expectation over the full support of $\mathcal{X}_0$. Additionally, we draw $M$ samples uniformly from the counter-example buffer $\mathcal{B}^q$.
The empirical objective is
\begin{align}
         \min_{\theta \in \Theta, \nu, \zeta} \quad & \mathcal{L}(\theta, \nu, \alpha, \{x_{0}^k\}_{k=1}^{N+M}, \varphi) + \lambda \|\zeta\|_{2}^{2} \label{eq:lossobjective}\\ 
         \text{s.t.} \quad & \dot{x} = f(x; \theta)  + \mathrm{NN}(x;\zeta), \quad x_{0} \in \mathcal{X}_0,
\end{align}
which we solve with \ac{sgd} for the parameters $\theta, \zeta$, and $\nu$.
Specifically, we augment $f$ with a residual \ac{node} $\mathrm{NN}(x;\zeta)$, where the input is the current state $x$, and the parameters of the neural network are denoted by $\zeta$, which the objective \eqref{eq:lossobjective} regularizes. Intuitively, the \ac{node} smooths the objective over the parameters $\theta$.
Ideally, the objective would be to maximize the minimal robustness observed over the domain $\mathcal{X}_0$.
For computational tractability, this is relaxed to basing the component $\mathcal{L}$ of \eqref{eq:lossobjective} on the Conditional Value at Risk (CVaR) objective proposed in~\cite{rockafellar2000}, referred to as SlackReLU:
%This resembles the CVaR objective presented in~\cite{rockafellar2000}.
\begin{align}\label{eq:slackrelu}
    \mathcal{L}&(\theta, \nu, \alpha, \{x_{0}^k\}_{k=1}^{N+M}, \varphi) =  \\
    &\frac{1}{N+M} \sum_{i=1}^{N+M} \left[ \max(0, \nu -\rho(\vb{x} (x_{0}^k, \theta), \varphi)) - \alpha \nu \right], \notag
\end{align}
where $\alpha$ is a hyperparameter. A small $\alpha$ is preferred to emulate the true objective, which is to maximize the worst-case robustness. This comes at the expense of training stability as a smaller $\alpha$ tracks gradients for very few samples.
$\mathcal{L}$ is minimized for the smallest $\nu$ such that a fraction $1-\alpha$ of the initial conditions have robustness exceeding $\nu$.  

\begin{proposition}[Approximate Soundness of SlackReLU]

If the optimal $\mathcal{L}^\star$ in \eqref{eq:lossobjective} is negative, then the average robustness of the worst $\alpha$ fraction of the samples $x_0^k$ is $\nu^\star > 0$, and the remaining $1 - \alpha$ fraction of the samples $x_0^k$ have robustness at least $\nu^\star$ everywhere.

\begin{proof}
    Directly follows from \eqref{eq:lossobjective} and \eqref{eq:slackrelu}, when $x_0^k$ are re-sampled every epoch.
\end{proof}
\end{proposition}

Alg.~\ref{alg:sgd} terminates after $E$ epochs, and selects the best parameters of the $q$ candidates according to
\begin{equation}
    \theta^* = \argmax_{\theta \in \{\theta^i\}} \underbrace{\min_{x_0^k \in X_\mathrm{test}} \rho(\vb{x}(x_0^k; \theta), \varphi)}_{\underline{\rho}(\theta, X_\mathrm{test})}.
\end{equation}
Note that \eqref{eq:discretetimeproblem} holds when $\underline{\rho}(\theta^*; X_\mathrm{test}) > 0$. \vspace{0.1cm}

\textbf{Continuous-domain Verification.}
To obtain the continuous-time guarantee \eqref{eq:conttimeproblem}, we use set-based reachability analysis for nonlinear systems \cite{Wetzlinger2021} to compute the reachable set $\mathcal{R}_{[0,T]}$ \eqref{eq:reachableset} of $f(x;\theta^\ast)$, and verify the \ac{stl} specification $\varphi$ using set-based \ac{stl} \cite{lercher24}. We employ \cite{Wetzlinger2021, lercher24} due to their computational efficiency and tight reachable sets. In general, our approach is agnostic to the reachability algorithm \cite{althoff2021set, Meyer2021} and set-based \ac{stl} \cite{roehm2016stl, Baird2023}.

\section{Evaluation Systems}

We demonstrate our method on three dynamical systems that differ in their \ac{ode} structure. They range from 6 to 12 state dimensions and 6 to 18 parameter dimensions. The first system is a quadrotor described by a 12-dimensional \ac{ode} model, for which we demonstrate the joint synthesis of system parameters and control gains.
The second system is a gene network with 18 parameters and a complex specification. 
The third system is the Laub-Loomis model of enzymatic activity, a verification benchmark \cite{geretti2025arch}.
In the following, we describe the parametric models, initial conditions, and \ac{stl} specifications for each system.

\subsection{PD-controlled Quadrotor}
We use a controlled quadrotor \cite[Chap. 8.2]{Meyer2021},
given by the second-order \ac{ode} for position and orientation domains:
\begin{equation}\label{eq:quadrotor}
\begin{aligned}
    \ddot{p_\mathrm{n}} &= \frac{F}{m} \left( -\cos(\phi) \sin(\chi) \cos(\psi) -  \sin(\phi)\sin(\psi) \right) \\
    \ddot{p_\mathrm{e}} &= \frac{F}{m} \left( -\cos(\phi) \sin(\chi) \sin(\psi) +  \sin(\phi)\cos(\psi) \right) \\
    \ddot{h} &= \frac{F}{m} \cos(\phi) \cos(\chi) - g \\  
    \ddot{\phi} &= \frac{1}{I_{x}} \tau_\phi ,\, \ddot{\chi} = \frac{1}{I_{y}} \tau_\chi ,\, \ddot{\psi} = \frac{1}{I_{z}} \tau_\psi, \quad %\text{where}
\end{aligned}
\end{equation}
where
\begin{equation}\label{eq:quadrotor_2}
\begin{aligned}
    F &= mg - k_p (h - h_{\mathrm{ref}}) - k_d (\dot{h}) \\ 
    \tau_\phi &= - \phi - \dot{\phi} ,\, \tau_\chi = - \chi - \dot{\chi}, \, \tau_\psi = - \psi - \dot{\psi}  \\ 
    m &= M_\mathrm{body} + 4  M_\mathrm{rotor} \\ 
    I_x &= I_y = \frac{2}{5} M_\mathrm{body} R^2 + 2  M_{\mathrm{rotor}} L^2 \\
    I_z &= \frac{2}{5} M_\mathrm{body} R^2 + 4 M_{\mathrm{rotor}} L^2,
\end{aligned}
\end{equation}
$g$ is $\SI{9.81}{\meter \per \second \squared}$ and $h_\mathrm{ref} = \SI{1}{\meter}$. The initial conditions are the hyperrectangle with $[-0.4, 0.4]$ for the cartesian positions $p_\mathrm{n}, p_\mathrm{e}, h$, and their velocities $\dot{p_\mathrm{n}}, \dot{p_\mathrm{e}}, \dot{h}$. The angles $\phi, \chi, \psi$ are set to $0$, and rotational velocities $\dot{\phi}, \dot{\chi}, \dot{\psi}$ are within $[-0.02, 0.02]$. The time horizon is $T=\SI{5}{\second}$ and trainable system parameters in \eqref{eq:quadrotor} are 
$\theta = \begin{bmatrix} M_\mathrm{body} & M_\mathrm{rotor}  & R   & L   & k_p   & k_d \end{bmatrix}^{\transp}$, and the actuation force $F$ is restricted to the range $[\SI{10}{N}, \SI{30}{N}]$ via $\tanh$.
As in \cite[Chap. 8.2]{Meyer2021}, we aim to verify that the quadrotor never exceeds the height of $\SI{1.4}{\meter}$, and stays above $\SI{0.9}{\meter}$ meters after the first second. We extend the specification to ensure that the vertical speed is smaller than $\SI{0.1}{\meter\per\second}$. Formally:
\begin{equation}\label{eq:quad-spec}
\scalemath{0.92}{
    \varphi: \; \always_{[1s, 5s]} (h \geq 0.9) \land \always(h \leq 1.4) \land  \always_{[3s,5s]}(| \dot{h}| \leq 0.1).
}
\end{equation}

\subsection{Signal-propagating Gene Regulatory Network}
The next system models signal propagation in gene networks, built on activation, inhibition, and decay terms~\cite{Krasowski2025a}: 
\begin{equation}\label{eq:hill}
    \begin{aligned}
        \dot{x}_1  &= \nu_1 \frac{1}{1 + k_{41}x_{4}^{2}}  - \gamma_1 x_1 + 0.5 \\
        \dot{x}_2 &= \nu_2 \frac{ k_{12} x_{1}^{2} + k_{42}x_{4}^{2} + k_{52}x_{5}^{2} }{1 +  k_{12} x_{1}^{2} + k_{42}x_{4}^{2} + k_{52}x_{5}^{2} }  - \gamma_2 x_2 + 0.5 \\
        \dot{x}_3 &= \nu_3 k_{53 } x_{5}^{2} - \gamma_3 x_3  \\
        \dot{x}_4 &= \nu_4 \frac{1}{1 + k_{34}x_{3}^{2}} - \gamma_4 x_4  \\
        \dot{x}_5  &=  - \gamma_5 x_5  \\
        \dot{x}_6 &= \nu_6 k_{56} x_{5}^{2}  - \gamma_6 x_6
    \end{aligned}
\end{equation}
The set of initial conditions is a hyperrectangle defined by the state lower bound $[0.0, 0.0, 0.0, 0.0, 0.9, 0.9]$ and upper bounds $[0.4, 0.4, 0.4, 0.4, 1.0, 1.0]$. The time horizon is $T=\SI{25}{\second}$ and the system has 18 parameters to synthesize, i.e., all $\gamma$, $\nu$, and $k$.
We verify the following specification, adapted from \cite{Krasowski2025a}, which enforces that $x_3$ and $x_4$ must initially grow.
However, when $x_4$ exceeds a certain level, $x_3$ must decay. 
% We aim to verify the following specification, adapted from \cite{Krasowski2025a}, which describes signal propagation and enforces $x_1 \geq 0.2$ and $x_2 \geq 0.3$:
\begin{equation}\label{eq:hill-spec}
    \begin{aligned}
        \varphi: \; & \event \left( x_1 \geq 0.2 \land x_2 \geq 0.3 \right) \\
                & \land \big( (x_1 \geq 0.2 \land x_2 \geq 0.3) \\ 
                & \;\;\implies (\event_{[0,10]} \,  x_3 \geq 0.5) \land (\event_{[0,10]} \always \,  x_4 \geq 0.9) \big)\\
            & \land \left( x_4 \geq 0.6 \implies \event_{[0,20]} \always \, x_3 \leq 0.3\right) \\
            & \land \left(\always x_1 \leq 1.5 \right)\land \left(\always x_2 \leq 1.5 \right) \\ 
            & \land \left(\always x_3 \leq 1.5 \right)\land \left(\always x_4 \leq 1.5 \right).
    \end{aligned}
\end{equation}

\begin{figure*}
    \centering
        \gradientbar[width=1cm, height=0.5mm, border=0mm, left label = SlackReLU]{mplC0} 
        \gradientbar[width=1cm, height=0.5mm, border=0mm, left label = ReLU]{mplC1} 
        \gradientbar[width=1cm, height=0.5mm, border=0mm, left label = LeakyReLU]{mplC2} 
        \\
    \begin{subfigure}[t]{0.32\linewidth}
        \centering
        \includegraphics[width=\linewidth]{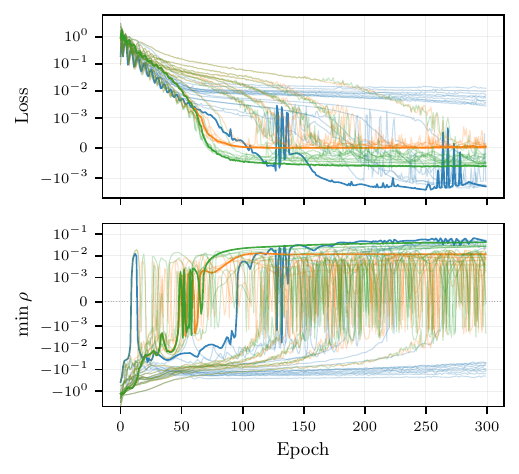}
        \caption{Quadrotor}
        \label{fig:quadcompare}
    \end{subfigure}
    \hfill
    \begin{subfigure}[t]{0.32\linewidth}
        \centering
        \includegraphics[width=\linewidth]{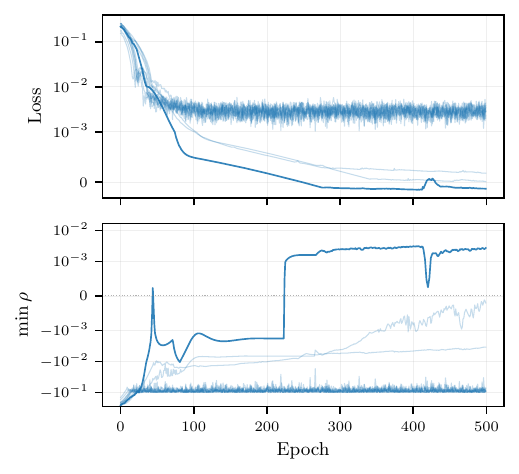}
        \caption{Gene network}
        \label{fig:hilltrain}
    \end{subfigure}
    \hfill
    \begin{subfigure}[t]{0.32\linewidth}
        \centering
        \includegraphics[width=\linewidth]{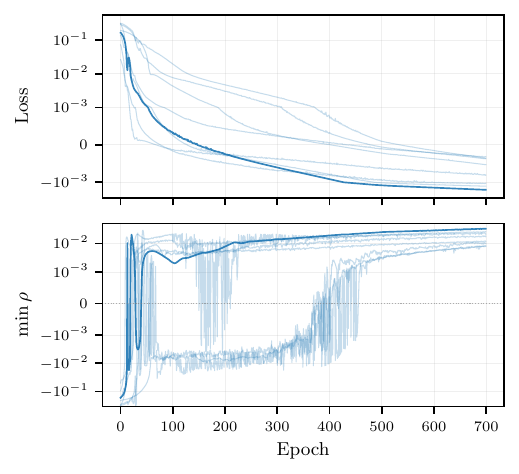}
        \caption{Laub Loomis}
        \label{fig:training_c}
    \end{subfigure}
    \caption{Training curves with best parameter highlighted for the objective function (top), and minimum robustness of each batch (bottom). (a) Includes a comparison of different loss functions.}
    \vspace{-0.2cm}
    \label{fig:training}
\end{figure*}

\subsection{Laub-Loomis Enzymatic Model}
The seven-dimensional Laub-Loomis model describes enzymatic activity and has been regularly used for the ARCH competition on system verification~\cite{geretti2025arch}: 
\begin{equation}\label{eq:laubloomis}
    \begin{aligned}
        \dot{x}_{1} &= k_1 x_3 - k_2 x_1 \quad        & &\dot{x}_{2} = k_3 x_5 - k_4 x_2 \\ 
        \dot{x}_{3} &= k_5 x_7 - k_6 x_2 x_3  \quad   & &\dot{x}_{4} = k_7 - k_8 x_3 x_4 \\ 
        \dot{x}_{5} &= k_9 x_1 - k_{10} x_4 x_5 \quad & &\dot{x}_{6} = k_{11} x_1 - k_{12} x_{6} \\
        \dot{x}_{7} &= k_{13} x_6 - k_{14} x_{2} x_{7}. \quad & &
    \end{aligned}
\end{equation}
The uncertain initial conditions are the same as those in the ARCH competition~\cite{geretti2025arch}: a hypercube with width $0.2$ around $x_{c} =  [1.2, 1.05, 1.5, 2.4, 1, 0.1,  0.45]$. We train the parameter vector $\theta = \begin{bmatrix} k_1 & \cdots  & k_{14}  \\ \end{bmatrix}$ and the time horizon is $T=\SI{20}{\second}$.
The seven-dimensional system has two stable equilibria; we design the specification to synthesize parameters for the non-zero equilibrium \cite{ghaemi2007}. 
%The other equilibrium has no species at $0$ if the $k_i$ are strictly positive.
We also aim to certify that $x_4$ stays below $4.5$ (which is falsified for the ARCH system):
\begin{align}
    \varphi: \; &\always_{[0,20]} \left(x_4 \leq 4.5 \right) \land \\
    %&
    % \textcolor{red}{\always_{[0, 16]}\left( (x_4 \geq 3.0) \implies \event_{[0,4]}(x_4 \leq 3.0)\right) \, \land } \notag\\
    % & \textcolor{blue}{\event_{[0,10]} \always \left( x_2 \geq 0.1 \right) \land \event_{[0,10]} \always \left( x_7 \geq 0.1 \right)} \notag\\
    %&\event_{[0,10]} \always \left( |x_3 - \textcolor{red}{C} x_1 | \leq 0.1 \right).\\
    &\event_{[0,10]} \always \left( |x_3 - \overline{x}_3 | \leq 0.1 \right) \land
     \event_{[0,10]} \always \left( |x_2 - \overline{x}_2 | \leq 0.1 \right).\notag
\end{align}
where $\bar{x}_{2}=0.35$ and $\bar{x}_{3} =0.55$ are approximately equilibria for the ARCH competition system~\cite{ghaemi2007, geretti2025arch}.

\section{Numerical Experiments}\label{sec:results}
We present the results for our approach for the three systems by answering the questions:
\begin{enumerate}
    \item Does our method robustly find satisfying parameters?
    \item How effective are the proposed relaxations?
    \item Can we provide continuous-time guarantees for parameters trained using discrete-time signals?
\end{enumerate}

%\textbf{Experiment setup. }
In all experiments, we ensure the non-negativity of parameters by treating the $\log$ of these parameters as trainable, and exponentiating for simulating the system.
The time step size for the \ac{ode} solver during training is $\SI{1}{\second}$.
%and the error tolerance thresholds are set to $10^{-4}$ and $10^{-5}$ for absolute and relative tolerances, respectively.
% We always train with $256$ samples in the interior of $\mathcal{X}_0$ and $64$ samples on the boundary for each epoch, i.e., $N=320$. For the high-dimensional initial condition rectangles, the first $n / 2$ are sampled from corners, then $n / 4$ are sampled from boundaries of codimension $1$, and so on until all codimensions are exhausted. 
The capacity of the first-in-first-out counter-example buffer $\mathcal{B}$ is $1024$, and at each epoch, we select $M=32$ uniform random samples from $\mathcal{B}$.
The residual network is a Multi-layer Perceptron (MLP) with two hidden layers of size 64, and regularizer $\lambda= 0.1$. The SlackReLU coefficient is set to $\alpha = 0.05$ and $R=128$.
For the quadrotor, gene network, and Laub-Loomis model, we train for $E= 300, 500$, and $700$ epochs with learning rates 0.05, 0.005, 0.002, respectively. We keep $q=16, 12$, and $8$ parameter initializations based on evaluating with $P =$ 262,144, 46,656, and 279,936 initial conditions, respectively.
% The parameter initialization distribution $\mathcal{T}$ of the quadrotor are log-normal distributions for $M_\mathrm{body}, M_\mathrm{rotor}, R$ and $L$. The gains $k_p, k_d$ are uniformly sampled from a best-guess range of $[0, 30]$ and $[0, 10]$ respectively.
% For the gene network, the logarithm of the trainable parameters are uniformly sampled from the interval $[0.001, 1]$. For the Laub Loomis model, parameters are drawn from a log-normal with mean $0$ and variance $1$. 
We use JAX~\cite{jax2018} for parallelization.\footnote{Code: \href{https://github.com/Beau-Coup/STLParameterSynthesis.git}{github.com/Beau-Coup/STLParameterSynthesis.git}}

\begin{figure}[b!] 
\centering
\vspace{-0.3cm}
    \includegraphics[width=0.4\textwidth]{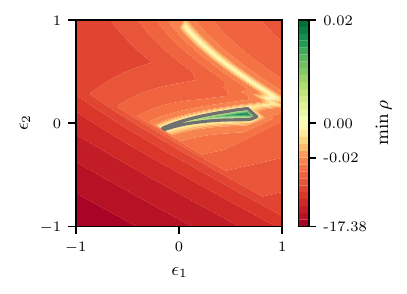}
    \vspace{-0.3cm}
    \caption{Loss landscape for the gene regulatory network. $\epsilon_1, \epsilon_2$ are random orthogonal unit vectors in parameter space.
        The origin is the optimal parameters found with Alg.~\ref{alg:sgd}.
        Random vectors provide an unbiased view of the landscape at that point.
        The gray curve shows the zero level contour.
    }
    \label{fig:landscape}
\end{figure}

\begin{figure}[b!]
    \begin{center}
        \vspace{-0.3cm}
        \includegraphics[width=0.4\textwidth]{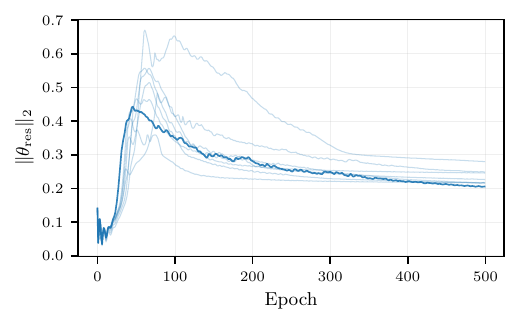}
        \vspace{-0.3cm}
        \caption{Norm of residual MLP over training for multiple parameter initializations for gene regulatory network.}
        \label{fig:hill-l2}
    \end{center}
\end{figure}

\subsection{Robust Learning}
To investigate training convergence, we show the loss and minimum robustness of the training batches for the three systems for the parameter initializations (also: parameter seeds) in Fig.~\ref{fig:training}. 
The curves clearly demonstrate that initialization with high robustness parameter candidates does not ensure convergence to satisfying parameters and often only a few parameter initializations reach satisfaction. 
We observed that this is mitigated by increasing the number of epochs by additional 200, e.g., 1 out of 12 seeds converge for the gene network and all converge for the Laub-Loomis model. %, which is trained for an additional 200 epochs.
% Note that we observed that the higher sensitivity of SlackReLU to initialization can be reduced by longer training, which is likely due to the increased capacity when adding $\nu$.

Additionally, for the quadrotor we compared the choice of using SlackReLU with LeakyReLU and ReLU (see Fig.~\ref{fig:quadcompare}). 
The training curves show that SlackReLU converges for fewer parameter initializations, but reaches the best minimum robustness. 
%Thus, striking the balance between pushing parameters beyond the satisfaction boundary (unlike ReLU), without over-optimizing for already satisfied initial conditions.
While LeakyReLU and SlackReLU appear competitive with respect to minimum robustness, the LeakyReLU and ReLU losses are significantly more volatile.

\begin{figure*}
    \centering
     \hspace{6cm}
    \gradientbar[font=\scriptsize,width=0.8cm, height=0.5mm, border=0pt, left label=$x_1$]{mplC0} \hspace{0.3em}
        \gradientbar[font=\scriptsize,width=0.8cm, height=0.5mm, border=0pt, left label=$x_2$]{mplC1} \hspace{0.3em}
        \gradientbar[font=\scriptsize,width=0.8cm, height=0.5mm, border=0pt, left label=$x_3$]{mplC2} 
        \gradientbar[font=\scriptsize,width=0.8cm, height=0.5mm, border=0pt, left label=$x_4$]{mplC3} \hspace{0.3em}
        \gradientbar[font=\scriptsize,width=0.8cm, height=0.5mm, border=0pt, left label=$x_5$]{mplC4} \hspace{0.3em}
        \gradientbar[font=\scriptsize,width=0.8cm, height=0.5mm, border=0pt, left label=$x_6$]{mplC5} \hspace{0.3em}
        \gradientbar[font=\scriptsize,width=0.8cm, height=0.5mm, border=0pt, left label=$x_7$]{mplC6}\\
    \begin{subfigure}[t]{0.32\linewidth}
        \centering
    %     \gradientbar[width=1.5cm, height=0.5mm, border=0pt, left label=0, right label=0.09]{mplC2} \\
    % \vspace{-1em}
    \includegraphics[width=\linewidth]{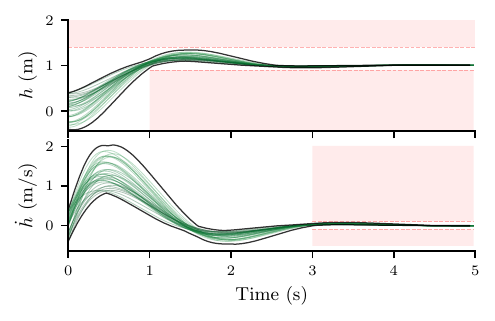}
        \caption{Quadrotor}
        \label{fig:quadtraj}
    \end{subfigure}
    \hfill
    \begin{subfigure}[t]{0.31\linewidth}
    \includegraphics[width=\linewidth]{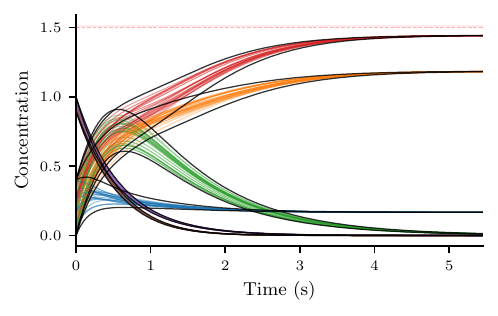}
        \caption{Gene network}
        \label{fig:hilltraj}
    \end{subfigure}
    \hfill
    \begin{subfigure}[t]{0.32\linewidth}
        \centering
        \includegraphics[width=\linewidth]{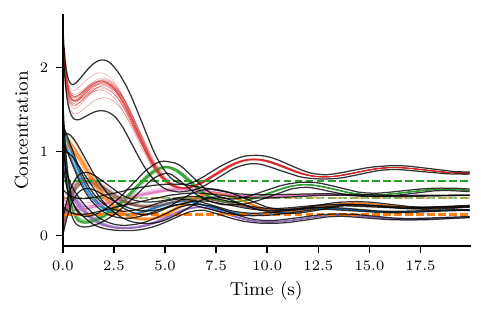}
        \caption{Laub Loomis}
        \label{fig:laub-verify}
    \end{subfigure}
    \caption{Trajectories for various initial conditions. More saturated trajectories indicate higher robustness. The outer black bands indicate the upper and lower bounds of the reachable sets given from the verification step. (a) Violation regions are in red. (b) The maximum concentration is indicated by the red line, and we truncate to \SI{5}{s}, as the system reaches steady state by then.}
    \label{fig:verification}
    \vspace{-0.2cm}
\end{figure*}

\subsection{Ablation of Relaxations}
%our failure sampling
We investigate the effects of the proposed problem relaxations, i.e., the parameter initialization and the residual network on training, for the gene network, which has the highest parameter dimensionality and the most complex specification.
In Fig.~\ref{fig:landscape}, we illustrate the loss landscape for the found optimal parametrization. 
%While the optimized parameter vector satisfies the spec, it is located on a steep hill in the loss landscape with a large flat area. 
We observe a large attractive region toward a small irregular local minimum,
%During optimization, there is no guarantee that the local minimum results in a vector with positive robustness.
which is due to the \ac{stl} robustness semantics relying on $\max$ and $\min$ operators, which render $\rho$ non-smooth. 
%This highlights the importance of simultaneous parameter optimization, especially since only one converges to satisfaction for the gene network.
Fig.~\ref{fig:hill-l2} shows the Euclidean norm of residual network parameters over training. The parameters initially increase to guide the \ac{ode} solution to a satisfactory assignment before gradually decreasing. The satisfying parameter seed results in the lowest number of residual network parameters.
%Without the residual network, learning a satisfying parameterization required many more parameter samples.

% \textcolor{red}{One challenge with ReLU is that it does not incentivize the parameters $\theta$ to cross the satisfaction boundary and achieve robustly positive robustness --- it only has gradients when the robustness is negative.
% There are many alternatives, such as LeakyRelu, which replaces the $0$ by a smaller linear term $-0.01 \cdot \rho$, or even smooth versions like SiLU, which smoothly transitions from $0$ to a linear function for negative $\rho$.
% We consider LeakyReLU in this paper, but omit smooth versions since smooth derivatives at $0$ dilute the priority of achieving positive robustness on all initial conditions, and muddy the soundness of the resulting loss.}

\subsection{Model-based Continuous-Time Verification}
For all optimal parameters found, we verified the continuous-time system by using set-based reachability analysis \cite{Wetzlinger2021} with the incremental reachset temporal logic \cite{lercher24} in CORA \cite{cora}. 
Fig.~\ref{fig:verification} illustrates the computed reachable sets for state dimensions relevant to the specification.
The verification for the Laub-Loomis model indicates that our algorithm effectively re-adjusts parameters if a specification is violated. Specifically, now the $4.5$ threshold for $x_4$ can be verified while reaching a non-zero equilibrium close to the equilibrium of the ARCH competition system, which violated this threshold~\cite{geretti2025arch}.

% \begin{figure}[h!]
%     \begin{center}
%         {\footnotesize Robustness}
%         \gradientbar[width=1.5cm, height=0.5mm, border=0mm, left label=0, right label=0.08]{mplGreen}
%         \includegraphics[width=0.5\textwidth]{figures/quad-traj-relu.pdf}
%         \caption{Quadrotor trajectories for various initial conditions from the ReLU loss.}
%         \label{fig:quadtrajrelu}
%     \end{center}
% \end{figure}

\section{Conclusion}
This paper presents a parameter synthesis method for parametric \ac{ode} using \ac{stl} specifications.
We demonstrate that the approach scales to high-dimensional systems, consistently identifies valid parameters, and provides strong continuous-time guarantees of specification satisfaction.
Moreover, it enables synthesizing system parameters from qualitative semantics of formal system requirements that are interpretable in natural language.

For our experiments, a time step size of $\SI{1}{\second}$ was sufficient to learn satisfying parameters. 
Coarser steps ease compliance with safety specifications but complicate liveness specifications. Finer time steps increase the computational burden of the gradient evaluations and \ac{ode} solver. Because the selected step size depends on the system dynamics and specifications, systematically analyzing these trade-offs remains an important future research direction.

% The evaluation systems demonstrate the feasibility and effectiveness of our approach across a range of systems solely based on diverse \ac{stl} specifications.
% However, gradient-based approaches are incredibly reliable for regressing on data.
% If available, incorporating data could potentially make the synthesis step more reliable and robustify the identified parameters with respect to the specification.
% Specifically, using candidate parameter vectors, real-world parameters for related systems, or sparse time series data presents an interesting avenue for future work.
% %Or, for example, generating realistic but more complex parameter sets $\Theta$ from data such as larger classes of compact sets than the ones considered here could be a significant improvement.

A main challenge for applying gradient-based methods on \ac{stl} is its non-smoothness. 
Smooth versions of \ac{stl} robustness have been proposed~\cite{leungBackpropagationSignalTemporal2023, haghighiControlSignalTemporal2019}.
Preliminary experiments indicated that simply smoothing the robustness function at best did not improve, and at worst hindered, parameter synthesis.
%On the other hand, it is reasonable to expect that a gradient-based method benefits from smooth \ac{stl} semantics. 
Additionally, smooth \ac{stl} versions often sacrifice soundness, motivating further research. %since such semantics can be leveraged for efficient parameter synthesis, control, or verification.


%%%%%%%%%%%%%%%%%%%%%%%%%%%%%%%%%%%%%%%%%%%%%%%%%%%%%%%%%%%%%%%%%%%%%%%%%%%%%%%%

%%%%%%%%%%%%%%%%%%%%%%%%%%%%%%%%%%%%%%%%%%%%%%%%%%%%%%%%%%%%%%%%%%%%%%%%%%%%%%%%



% \section*{Acknowledgment}
% This work was funded in part by the Air Force Office of Scientific Research grant FA5590-23-1-0529.
% Alex Beaudin is partially supported by a Fonds de Recherche du Qu\'ebec Doctoral Scholarship.
%\textcolor{red}{TBD - NSF CPS Frontier? MURI}

%%%%%%%%%%%%%%%%%%%%%%%%%%%%%%%%%%%%%%%%%%%%%%%%%%%%%%%%%%%%%%%%%%%%%%%%%%%%%%%%
\bibliography{literature, sources}
\bibliographystyle{IEEEtran}

\end{document}